\documentclass[11pt,twoside]{article}

\usepackage{asp2006}
\usepackage{lscape}
\usepackage{graphicx}

\begin{document}
\title{Dark Matter Substructure in Lensing Galaxies}
\author{Masashi~Chiba,$^1$ Takeo~Minezaki,$^2$ Kaiki~T.~Inoue,$^3$
Nobunari~Kashikawa,$^4$ Hirokazu~Kataza,$^5$ Hajime~Sugai$^6$}
\affil{$^1$Astronomical Institute, Tohoku University, Sendai, Japan\\
$^2$Institute of Astronomy, University of Tokyo, Mitaka, Tokyo, Japan\\
$^3$School of Science and Engineering, Kinki Univ., Higashi Osaka, Japan\\
$^4$National Astronomical Observatory of Japan, Mitaka, Tokyo, Japan\\
$^5$ISAS, JAXA, Sagamihara, Kanagawa, Japan\\
$^6$Department of Astronomy, Kyoto University, Kyoto, Japan}

\begin{abstract}
To set useful limits on the abundance of small-scale dark matter halos
(subhalos) in a galaxy scale, we have carried out mid-infrared imaging and
integral-field spectroscopy for a sample of quadruple lens systems
showing anomalous flux ratios. These observations using Subaru
have been successful for distinguishing millilensing by subhalos
from microlensing by stars. Current status for our lensing analysis
of dark matter substructure is reported.
\end{abstract}

%% Section 1 %%
\section{Introduction}

Recent high-resolution N-body simulations based on Cold Dark Matter
(CDM) theory highlight a so-called missing satellite problem, i.e.,
CDM predicts the existence of more than several hundred dark satellites
(or subhalos) in a galaxy-sized halo, in sharp contrast to
the observed number of about 20 Milky Way satellites
(e.g., Diemand et al.~2007 for recent studies). 
To clarify this issue, gravitational lensing offers us an invaluable insight
into such numerous CDM subhalos that reside in lensing galaxies.
In particular, anomalous flux ratios in lensed QSOs, namely those
hardly reproduced by any lens models with a smooth density distribution,
are of special interest, since lens substructures are able to cause
such flux anomalies (e.g., Metcalf \& Madau 2001; Chiba 2002).

Here, we report on our Subaru observations of a sample
of quadruple lenses with anomalous flux ratios, based on mid-infrared
imaging and integral-field spectroscopy.
Mid-infrared imaging of lenses is advantageous because the
flux is free from differential extinction among different images and
it is also free from microlensing by stars. Integral-field spectroscopy
provides both spatial and spectral information simultaneously
on each lensed image.
We select eight QSOs with four lensed images, PG1115$+$080,
B1422$+$231, MG0414$+$0534, Q2237$+$030, H1413$+$117, HS0810$+$2554,
and WFI2026$-$4536 for mid-infrared imaging, and RXJ1131$-$1231 for
integral-field spectroscopy. We briefly report the results for
PG1115$+$080 and B1422$+$231 (Chiba et al. 2005) and RXJ1131$-$1231
(Sugai et al. 2007); those for other targets will be published
elsewhere (Minezaki et al. 2008 in preparation).
We present new limits on substructure in the observed
lensed systems and implications for a missing satellite problem are discussed.

%% Section 2 %%
\section{Targets and Observations}

PG1115$+$080 at redshift $z_{\rm S}=1.72$ and B1422$+$231 at $z_{\rm S}=3.62$
are lensed by foreground ellipticals at $z_{\rm L} = 0.31$ and 0.34, respectively.
The former lens system holds the closely separated pair of images A1
and A2 with a separation of $0.\arcsec 48$,
and this configuration emerges if the QSO is close to and inside
a fold caustic provided by the lens (Figure 1). The latter shows
the colinear, three highly magnified images, A, B, and C, and this configuration
emerges if the QSO is close to and inside a cusp caustic. In such lens
systems associated with a fold or cusp caustic, there exists a
universal relation between the image fluxes, i.e., A2$/$A1$=1$
or (A$+$C)$/$B$=1$, whereas the observed optical flux ratios violate these rules
significantly, A2$/$A1$=0.64 \pm 0.02$ and (A$+$C)$/$B$ = 1.50 \pm 0.01$.

The mid-infrared imaging of PG1115$+$080 and B1422$+$231 was carried out
on the nights of UT 2004 May 5 and 6, using the cooled mid-infrared camera
and spectrometer (COMICS).
The field of view is $42\arcsec \times 32\arcsec $ and
the pixel scale is $0.\arcsec 129$ pixel${}^{-1}$.
We used the N11.7 filter, whose effective wavelength and bandwidth
are $\lambda_c = 11.67\ \mu$m and $\Delta \lambda=1.05\ \mu$m, respectively.
The FWHM of PSF was $0.\arcsec 33$ at small airmass.
The observed mid-infrared waveband
corresponds to the near-infrared waveband in the rest frame,
and its flux is dominated by thermal radiation
from hot dust located at the innermost region of a dust torus.
The inner radius of a dust torus, which is determined
by the highest sublimation temperature of dust ($T\sim 1800~K$)
and the UV luminosity of a QSO central engine, is generally much larger
than Einstein radii of foreground stars, so that
the observed mid-infrared flux is free from microlensing effects.

RXJ1131$-$1231 is unique in its low redshift of a source image
$z_{\rm S}=0.658$ lensed by an elliptical at $z_{\rm L}=0.295$.
The lens shows three roughly co-linear images,
A, B, and, C, being characteristic of a cusp singularity, where A is
brightest (Figure 2). The observed flux ratios, (B$+$C)$/$A$ \simeq 2.1$
in the $V$ band and $2.2$ in the $R$ band, deviate significantly
from the rule (B$+$C)$/$A$=1$.

Using the IFS mode of the Kyoto tridimensional spectrograph II (Kyoto 3DII),
we observed  RXJ1131$-$1231 on the night of UT 2005 February 8.
The IFS mode uses an array of $37 \times 37$ lenslets, enabling us
to obtain spectra of $\sim 10^3$ spatial elements. The spectral range from
$7300 \AA$ to $9150 \AA$ was observed in each of two one-hour exposures.
With the spatial sampling of $0^{\prime\prime}.096$ lenslet$^{-1}$,
the field of view of $\sim 3^{\prime\prime}$ covered the three bright
lensed images.
We measured the emission-line fluxes of
both the BLR H$\beta$ and
the NLR [OIII]$\lambda\lambda$4959,5007 for images
A, B, and C simultaneously.
%The measurements of the line fluxes are more
%reliable than those of the continuum because there are no contributions
%from the lensing galaxy, the redshift of which differs from the quasar's.
The H$\beta$ and [OIII] lines are very close in wavelength, so that
the effect of differential reddening between them is negligible.

%% Section 3 %%
\section{Results}

\begin{figure}[ht!]
\centering
\includegraphics[height=9cm]{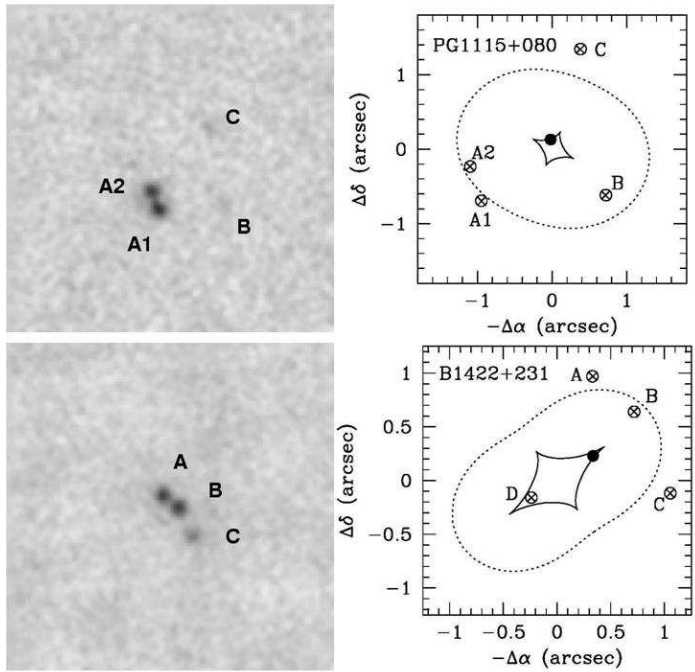}
\caption{Left: Mid-infrared images at $11.7\ \mu$m for PG1115$+$080 (top)
and B1422$+$231 (bottom). The north is up and the east is left.
Right: Smooth lens models for these systems, where solid and dotted
lines denote the caustics and critical curves, respectively,
and filled circles indicate the source positions.}
\end{figure}

The mid-infrared images at $\lambda = 11.7\ \mu$m of PG1115$+$080 and
B1422$+$231 are presented in Figure 1. It is evident that
the lensed images in concern, (A1, A2) and (A, B, C), respectively,
are clearly detected and well separated from each other.
%The minimum image separations are $0.\arcsec 48$ and $0.\arcsec 50$,
%respectively. 
%The faint lensed images B and C of PG1115$+$080 were also detected.
The mid-infrared flux ratios are given as
A2$/$A1$=0.93\pm0.06$, B$/$A1$=0.16\pm0.07$, and C$/$A1$=0.21\pm0.04$
for PG1115$+$080, and 
(A$+$C$)/$B$=1.51\pm0.06$, A$/$B$=0.94\pm0.05$, and C$/$B$=0.57\pm0.06$
for B1422$+$231.
This suggests that
the mid-infrared flux ratio for A2$/$A1 of PG1115$+$080
is consistent with the prediction of a smooth lens model ($\simeq 1$),
while the optical flux ratio is much smaller ($\simeq 0.65$), and that
the mid-infrared flux ratios for \rm{(A, B, C)} images of B1422$+$231
remain anomalous, contrary to the prediction of a smooth lens.

Based on the dust reverberation method for estimating the size of a dust
torus (Minezaki et al. 2004), we obtain an angular size $\theta_{\rm S}$
of a source image as $\theta_{\rm S} \simeq 1 \times 10^{-4}$ arcsec and
$3.7 \times 10^{-4}$ arcsec for PG1115$+$080 and B1422$+$231, respectively.
For PG1115$+$080, any substructure causing its optical anomalous flux ratio
A2$/$A1 should have a small Einstein angle $\theta_{\rm E}$ compared to
$\theta_{\rm S}$, because its mid-infrared ratio remains unaffected.
This suggests that a substructure mass inside $\theta_{\rm E}$,
denoted as $M_{\rm E}$, should be
smaller than 20~M$_\odot$, being comparable to the mass of a star,
i.e., microlensing causes an optical anomalous flux ratio.
For B1422$+$231, the presence of flux anomaly even
in mid-infrared waveband suggests $M_{\rm E} \ga 200$ M$_\odot$, i.e.,
lensing by a subhalo is most likely.

\begin{figure}[ht!]
\centering
\includegraphics[height=6cm]{chiba_fig2a.eps}
\includegraphics[height=4cm]{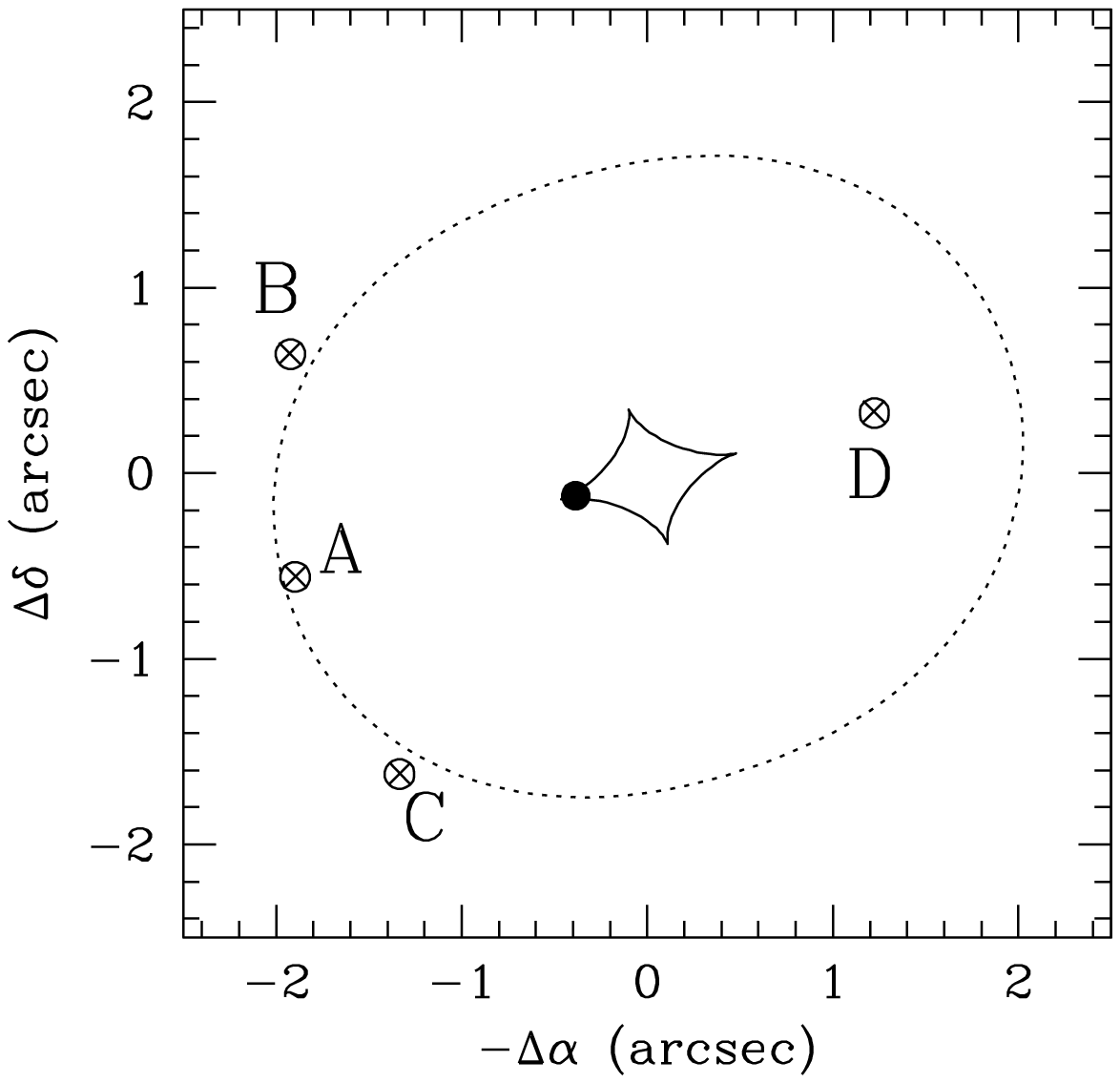}
\caption{Left: Kyoto 3DII spectra of RXJ1131$-$1231 for image A (solid line),
B (dashed line), and C (dot-dashed line), where
each has been extracted with an 8-lenslet, i.e.,
a circular aperture with a diameter $0^{\prime\prime}.77$.
Right: Smooth lens model for this lens system.}
\end{figure}

Figure 2 shows the Kyoto 3DII spectra of RXJ1131$-$1231 for
three bright images A, B, and C.
It is interesting to remark that images B and C in [OIII] show nearly
the same fluxes relative to image A, as expected from
a smooth model at its cusp singularity. Thus,
the absence of substructure lensing effects
on this NLR [OIII] sets important
limits on the mass of any substructures along the line of sight, 
as $M_{\rm E} < 10^5$ M$_\odot$.
In contrast, the H$\beta$ line emission, which originates from
the BLR, shows an anomaly in the flux ratio between
images B and C, i.e., a factor two smaller 
C$/$B ratio than predicted by smooth-lens models.
The ratio of A$/$B in the H$\beta$ line is well reproduced. 
The anomalous C$/$B ratio for the H$\beta$ line is caused most likely
by microlensing of image C, with $M_{\rm E} \ga 0.1$ M$_\odot$
for the mass of a substructure near image C.
We have also found the slight difference of the H$\beta$ line profile
in image A from those in the other images, which
suggests the presence of a small microlensing effect on image A.

%% Section 4 %%
\section{Prospects}

In addition to the above lens systems, we have already observed, using
COMICS, MG0414$+$0534, Q2237$+$030, H1413$+$117, and HS0810$+$2554,
and the calibration and analysis are underway. Our final target for mid-infrared
imaging of quadruple lenses, WFI2026$-$4536, is scheduled for observation
using Gemini-South this year.
Our preliminary statistical model implies that about 30~\% of the lens systems
($2-3$ out of 8) show a flux anomaly with $> 25$~\%, if subhalos are modeled by
tidally-truncated singular isothermal spheres with a mass function predicted
by the N-body simulations.
Once the concrete observational information for all of these targets
is ready, we will be able to set a more reliable
constraint on the abundance of CDM subhalos.

%\acknowledgements

\end{document}